\documentstyle[aps]{revtex}
\begin{document}
\draft


\title{A New Regime for Dense String Networks in the One Scale Model 
with Friction}

\author{
Charanjit S. Aulakh$^{(1)}$, Michiyasu Nagasawa$^{(2)}$  and
Vikram Soni$^{(3)}$}

\address{$^{(1)}$ {\it Dept. of Physics, Panjab University,
Chandigarh, India}}
\address{$^{(2)}${\it Dept. of Information Science, Faculty of
Science, Kanagawa University, Kanagawa 259-1293, Japan}}
\address{$^{(3)}$ {\it National Physical Laboratory, New Delhi, India}}

\maketitle

\begin{abstract}

We examine the compatibility between the one scale model with
friction for the evolution of cosmic string networks and the Kibble
mechanism for string network formation during a second order phase
transition.  We find a regime which connects (in a dramatically
short time $\sim .1 t_c$.) the dense string
network (small network scale $L\sim 1/T_c$ ) created by the
Kibble {\it{mechanism}} to the (dilute) Kibble {\it{regime}}
($L\sim t^{5/4}$) in which friction dominated strings remain
till times $t_* \sim (M_P/T_c)^2 t_c$.
The enormous loss of string length implied by this result, for
strings formed below the GUT scale,
is due to the continued dissipative motion of the string network
which follows from  the one scale model's
identification of the typical curvature radius with the network
scale and the fact that  the phase transition time $t_c\sim
M_{Planck}/T_c^2$
is much larger than the damping time scale $l_f\sim T_c^2/T^3$.
The significant implications for string mediated Baryogenisis
are also discussed briefly.

\end{abstract}

\vspace{0.3cm}

\vskip .2 true cm

\section{Introduction} 

Topological defects are a natural consequence of the symmetry
breaking phase transitions predicted by spontaneously broken
gauge theories of particle physics.
 They are produced in the early universe and can have
 important consequences for the structure of the present day universe
 \cite{topdef}.

Recently it has been  suggested that the baryon asymmetry
in our universe can be generated  by strings carrying
 electroweak flux \cite{vach,soni,brand,nagaconf}. Such scenarios have 
the
 advantage that they can automatically satisfy the out of equilibrium
 condition even though the electroweak phase transition is of second 
order.
 The amplitude of the baryon asymmetry generated
is directly dependent  on the string density  that survives
 after the phase transition.
  Since GUT scale cosmic strings can serve  as seeds
  for the formation of galaxies, the evolution of such string networks
has been much investigated both analytically and numerically. As we 
discuss
 below, the role of friction due to passage of  strings through
 the ambient plasma becomes increasingly important as we go to lower 
scales
 of string formation. For GUT scale strings, therefore,  friction plays
only a   limited role. It is perhaps for this reason that the evolution
 of string networks in the presence of friction has  scarcely been
 addressed \cite{topdef,kibble} . Recently ,however, an
 analytic model for string evolution with friction which includes the
 effects of velocity evolution has been analysed \cite{ms1,ms2}. This
 analysis has made it clear that there are notable differences
 in the evolution of strings in the presence of friction,
 which we now recount\cite{topdef,kibble}.

The frictional force per unit length experienced by the string network
due to its scattering of the particles of the the ambient plasma has
the generic form ${\vec F_f} =-\beta \gamma {\vec v}T^3$. Here $\vec 
v$  is
the  velocity of the string element relative to the plasma ,
$\gamma=(1-v^2)^{-1/2}$ , T is the temperature and $\beta$ a numerical
factor $\sim 1 - 10^{-3}$. This frictional force modifies the
equation of  motion of
the string in the expanding universe by replacing the Hubble
damping term coefficient
$2 {\dot{a}}/ a$ by ${2 {\dot{a}}/ a} + \beta T^ 3 a/\eta^2 $, where
the dot signifies a  conformal time derivative , $a$ is the cosmic scale
factor and $\eta$ is the string scale (vev). From the ratio of these
 terms it follows that  for
$T\geq \eta^2/m_P$ the frictional force completely dominates the Hubble
term and controls string evolution. This domination  extends upto a time
$t_*=\sigma^2 t_c$ , where $\sigma=m_P/\eta$ and $t_c$ is the time 
of
string formation. For strings created below $10^{6}$ GeV the friction
dominated evolution thus persists till the epoch of of matter domination 
,
whereas for GUT scale strings friction is clearly much less significant.

The length scale  characterizing frictional phenomena is
$l_{fr}\sim \eta^2/\beta T^3$.
Thus at the time $t_c$ of string formation one has
$l_{fr}/t_c \sim 1/(\sigma\beta)$.
This simple consideration ( that
the cosmological time scale $t_c$ is much larger than the
friction time scale) makes it clear that frictional phenomena
 have ample scope to modify the network even via powerlaw evolution of 
the
 network correlation length when , as is generically true for non-GUT
strings, $\sigma $ is a large number (and $\beta$ is not
pathologically small). This observation is the basis of
 the new transient regime uncovered in this paper.

Finally an important consequence of friction is that it prevents the
uncontrollable generation of small scale structure  that
characterizes undamped string evolution. Freely moving
strings inevitably exhibit cusping behaviour at some points on the 
string
where the velocity is close to luminal\cite{topdef}.
  This feature bedevils the study of
GUT scale strings since it prevents the string network from settling 
into a
scaling regime where the network is characterized by a single 
correlation
length that scales with the horizon. The continued importance of
motion on the smallest length scales necessitates the use of complicated
multiscale models \cite{kope}. In contrast , since friction
ensures that the velocity of the string is always far from
luminal, small scale structure generation is unlikely to be
important in the friction dominated regime.
Thus there is good reason to believe \cite{ms1,ms2}
that even a simple single scale model can capture
the gross features of the friction dominated evolution, even at the high
initial densities predicted by the Kibble mechanism (see below),
provided string velocity evolution is  accounted for.

In the  commonly accepted
picture of string formation suggested by Kibble\cite{kibme}
$\epsilon_K$ string lengths of size $1/T_c$ are formed in each
correlation volume $\sim T_c^{-3}$ (where $\epsilon_K$ is a
probability factor $\sim 10^{-1} - 10^{-3}$) . This gives the
typical scale characterizing the network at formation as
$L_i\sim \epsilon_K^{-1/2}/T_c$.

Let us now turn to the details of the one scale model with friction 
within
whose framework we shall conduct our analysis of the the regime that
connects the state of the string network at formation (as predicted by 
the
Kibble mechanism) to the characteristic friction dominated regime , also
named after Kibble, where the string network is characterized by a 
terminal
velocity due to a balance between frictional and curvature forces.

As its name implies, the one scale model \cite{topdef}
 attempts to describe the
gross features of the string network in terms of a single length
scale $L$, whose evolution is determined by the values of a few
parameters only. Due to this  virtue of simplicity it
has assumed a paradigmatic value in the study of string networks.

  Since the frictional force is velocity
dependent it is clear that it's effect can be accounted for
correctly only by improving the one-scale model to include the
velocity evolution of the string network along with friction.
This improvement of the one scale model has recently been
presented in \cite{ms1}. The damped Nambu-Goto
equations for a friction dominated string {\cite{topdef}} are derived
 by  averaging
over the network and making the  assumption that the network curvature
scale $R$ is the same as the network correlation length $L$ at
all times.

 The analysis of \cite{ms1} , however,
does not cover the case of initial conditions appropriate to the
formation of string networks in second order phase transitions
via the Kibble mechanism \cite{kibme}. It is this lacuna that we seek to
 fill  in this letter.

In the one-scale model the (long) string length
density $n_L$ is by definition $L^{-2}$ where $L$ is the scale
which characterizes both the density and the average curvature
of the network .
The network evolution in a FRW background
is described by the following coupled first order differential
equations\cite{topdef,ms1,ms2}

\begin{eqnarray}
{dL\over dt}=& HL(1 + v^2) + \displaystyle{\frac{L v^2}{2l_f}}
+\displaystyle{\frac{c v}{2}}\ , \nonumber \\
{{dv \over dt}}=& (1-v^2)\left[ \displaystyle{\frac{k}{L}}-v\left( 2 H 
+
\displaystyle{\frac{1}{l_f}}\right)\right]\ ,
\label{lvt}
\end{eqnarray}
where $H,l_f$ are the Hubble parameter and friction length scale
respectively.

The friction length $l_f$ for gauge strings interacting with the cosmic
plasma  is given by $\eta^2/\beta T^3$ where $\beta$ is typically $\sim 
1$.
In the derivation of {\cite{ms1,ms2}}
the parameter $k$ characterizes the ``wiggliness'' of the
strings and can be expected to be $\sim 1$ when the network
motion is well damped by friction for the reasons discussed above.
 Thus $k\sim 1$ in the epoch when friction
is important while it can become small at the late times of the
linear scaling regime when
the strings are undamped and therefore  kinky .

 The parameter $c$ characterizes the rate at
which the network chops itself into loops by self-intersection.
For dense networks reconnection of loops back onto the long
string network may be important and we envisage that this effect
can at least qualitatively be captured by a lower effective
value of $c$. Such reconnection
processes have been observed in numerical
simulations \cite{abl} albeit in the case of semi-local strings.
Since we consider generic strings that do
intercommute rather than the special sorts that can only self
entangle we shall take $c$  $\sim 1 $ as well.

The analysis of \cite{ms1} uncovers the existence of three sequential
regimes : a ``stretching'' regime  ($L \sim L_i (t/t_c)^{1/2},
v \sim t (L_i\sigma)^{-1}$)  with very low velocities,
followed by the friction dominated or
``Kibble '' regime where $L\sim (t^5/t_c \sigma^{2})^{1/4},
v\sim (t/t_c \sigma^2)^{1/4}$
and finally the well known linear scaling regime where friction
 is unimportant and  $L$ scales
with the horizon while $v$ is essentially fixed and quasi
relativistic. The duration of the stretching regime
becomes shorter as the starting value $L_i$ is lowered (i.e for
denser networks). Indeed once $L_i\sim t_c/{\sqrt \sigma}$ the
simulations of \cite{ms1,ms2} show that  the
evolution appears to begin at $t_c$ with the Kibble regime itself,
omitting the stretching stage.
The initial value of $L$ in the Kibble regime is clearly
$>t_c/{\sqrt \sigma}$. 

The incompatibility of the Kibble mechanism and the Kibble
regime is now evident from the fact that
 the typical value of
$ L_i$ dictated by the Kibble mechanism
is $\sim t_c\epsilon_K^{-1/2} \sigma^{-1} $ which
is smaller than the  value at the onset of the Kibble regime
by a factor of $\sigma^{-1/2}$.
Thus, for typical values of $k,c \sim 1$,
there appears to be no regime which carries the network from
the dense state predicted by the Kibble Mechanism of string
formation  to the (dilute ) initial
states implied by the Kibble regime.

\vskip .5 true cm

{\section{New Transient Regime}}

As explained above the difficulty (and therefore its resolution)
lies in the short period between the phase transition and the
onset of the Kibble regime.  Consider the evolution of $L$ and $v$ in 
the
period immediately following the phase transition i.e when
${\hat t} = t-t_c' << t_c' ( or  t_c )$ , where $t_c'\sim  10t_c$ (say). The
crucial observation is that at this time the friction time scale
$l_f$ is smaller than the cosmological time scale by a factor
$\sim \sigma$. Thus a small cosmological time interval contains
many friction time periods so that friction can be very
effective in a ``short'' time.
The existence of the fast decay regime that emerges from our analysis
is not affected by the precise value of the of $t_c'\sim t_c$ at which 
the
evolution of the one scale model is taken to begin. This can be checked
by taking $t_c'$ to be any other multiple of $t_c$ in the range (say)
$t_c\leq t_c'\leq 100 t_c$ . The natural
length scale initially is $\eta\sim T_c$ so we scale $\hat t$ and L
by this length to define a dimensionless time $x=\eta \hat t$
and network scale $l=\eta L$. Note that this amounts to
measuring time in units of the damping time scale and and avoids
the introduction of large numbers in the evolution
equations which result if one uses $t_c$ as the unit of time. Then the
condition  ${\hat t} = t-t_c' << t_c$ translates to $x<< \sigma$.
 Since even for GUT scale strings $\sigma$ is $\sim
10^4$ the variable $x$ can change by many orders of magnitude before
the approximation becomes invalid. Expanding the terms in the
evolution equation around $t_c'$ one finds that $HL\sim l/\sigma$
while the friction term is $\sim l v^2$ and the chopping term is
$\sim v$. Thus as long as $l<\sigma^{1/2}$ and $v > \sigma^{-1/2}$
the Hubble term is completely dominated by the velocity
dependent terms. The evolution equations to leading order in
are simply

\begin{eqnarray}
2 { dl\over dx}=&  {{l v^2 \beta}} + c v\ , \nonumber \\
{dv\over dx}=& (1-v^2) \left( \displaystyle{\frac{k}{l}}
-{v \beta} \right)\ .
\label{lvx}
\end{eqnarray}

We wish to find the behaviour of $l$ and $v$
for large $x$ beginning from  natural initial conditions
(Kibble Mechanism) where
$l_i\sim \epsilon_K^{-1/2}$ .
>From the structure of the equation for $l$ it is clear that $l$
always increases . While $v$ will increase when $k\geq v\beta l$
and decrease in the opposite case. So if v is initially very
small or zero it will increase rapidly and then again fall off as
$l$ increases; while if it is $\sim 1$ then it will rapidly
decrease. Regardless of the initial value of $v$ the evolution
locks on  to a trajectory where $v$ is very
closely approximated by $k/{\beta l}$. This allows one to solve the
equation for $l$ to obtain the power law behaviours

\begin{eqnarray}
l=& \left( \displaystyle{\frac{(k +c) k}{\beta}}\right)^{1/2}x^{1/2}\ 
,
\nonumber \\
v=& \left(\displaystyle{\frac{k}{\beta (k+c)}}\right)^{1/2} x^{-1/2}\ 
.
\label{lv}
\end{eqnarray}

It is easy to check by substitution into eqn(\ref{lvt}) that this
regime persists until the growth of the Hubble term and the
decrease of the velocity dependent term makes them comparable
i.e till  $x\sim 10^{-1} \sigma$ (for typical values of $k,c$)
 by when $t\sim  t_c' + .1 t_c$.
After this the effect of the Hubble terms causes a shift in the
power-law exponents and the Kibble regime regime begins.  Note that
this is  compatible with the results of \cite{ms1,ms2}
who found that for higher initial network densities (i.e as they
varied  ${\tilde L_i}=L_i/t_c$  from $1$
 to $\sigma^{-1/2}$ -  the minimum initial network scale
considered by them) their ``stretching regime'' became shorter and
shorter till at the highest densities they considered, the Kibble
regime replaced the stretching regime to become the initial regime.
Since, however, the minimum initial network
length scale they considered was still
$\sigma^{1/2}$ times larger than the characteristic scale of a network
formed via the Kibble mechanism i.e $L \sim 1/\eta \sim 1/T_c$
a regime connecting these two very different scales was needed. The new
regime we have pointed out precisely fills the gap.

The most remarkable feature of this regime is thus that {\it the
values of $L$ and $v $ very shortly after the phase transition
are essentially independent of their initial values}. This
is in sharp contrast to the ``stretching regime '' $L\sim
L_c (t/t_c)^{1/2} ,
v\sim {t (L_c\sigma)^{-1}}$ found in
\cite{ms1,ms2} for initial conditions $\tilde L_i > \sigma^{-1/2}$.

It is easy to check that the Lyapunov exponents obtained by
linearizing the eqns(\ref{lvx}) around the solutions eqns(\ref{lv})
are indeed negative so that deviations from these solutions die
out exponentially fast as $x$ increases.
 We have verified the analysis given above by numerically
integrating the equations (\ref{lvx}) for various initial values
of $l,v$ at various values of the parameters $k,c,\sigma$ .
For large $x $ the asymptotic solutions
given in eqn.(\ref{lv}) are in excellent agreement with the
results of numerical integration of eqns.(\ref{lvx}) irrespective
of whether the initial velocities are relativistic or very small.

In the Kibble regime the network variables scale as follows
(${\tilde L} =L t_c^{-1}$):

\begin{eqnarray}
{\tilde L} =& \left(\displaystyle{\frac{2 k (k +c) }{ 3 \sigma}}
\right) ^{1/2} \left(\displaystyle{\frac{t}{t_c}}\right)^{5/4}\ , 
\nonumber
\\
v=& \left(\displaystyle{\frac{3 k }{2 (k+c) \sigma}}\right) ^{1/2}
\left(\displaystyle{\frac{t}{t_c}}\right)^{1/4}\ .
\label{lvkibb}
\end{eqnarray}

Matching the onset of the Kibble regime with the end of the
stretching regime and imposing the consistency condition that
this occurs after the phase transition shows that the stretching
regime can occur only if $\tilde L_i > \sigma^{-1/2}$. On the
other hand for networks arising from the Kibble mechanism
 ,where $\tilde L_i <<\sigma^{-1/2}$ ,
  the new transient regime neatly resolves
the inconsistency  by rapidly taking the network
from the dense state in which it is created to the dilute state
consistent with the fact that the minimum value of
$\tilde L_i$ compatible with the Kibble regime is $ \sim
\sigma^{-1/2}$ (for typical $k,c$).

This regime implies a very considerable loss of string length in
a very short cosmological time interval $\Delta t<<t_c'$. The
physical reason for this strange phenomenon is simply that near
the phase transition time the friction time scale
$l_f=\eta^2\beta/T_c^3\sim
1/T_c$ is much smaller than the cosmological time $t_c\sim
\sigma /T_c$ (for friction dominated gauge strings for which
$\sigma>>1$). On the other hand the one scale model identifies
the average network curvature with the network
correlation length and thus enforces curvature driven motion and
therefore dissipation.
 Since the regime  takes place over a time
interval during which the cosmological scale factor
changes very little the energy that is dissipated must come from
the string length : which in fact decreases sharply.
In \cite{ms2}the instructive example of
the damped motion of a circular string loop is discussed.
Such a loop can contract in radius and thus lose string length
even while preserving the transversal motion constraint
$ {\dot{\vec{x}}}\cdot {\vec{x}}'=0$ . The total
frictional work is exactly the energy loss due to decrease in
invariant string length for the loop due to its contraction.
Thus one may visualize the growth of the network correlation
length and the average curvature radius in tandem as being due
to the decaying away of curved and looped regions with high curvature
leaving behind a network with less string (larger $L$) and lower
curvature (larger $R$).

It is perhaps pertinent to point out that this analysis shows that friction
is by itself a genuine decay mechanism (independent of any specific
specific interaction like gravitational or electroweak radiation ) for both
the network and the loops.

\vskip .4 true cm

{\section {Discussion}}

The new regime presented here leads us to the conclusion that the one
scale model with friction presents a  consistent picture of network
evolution for initial network densities and velocities 
 given  by the Kibble mechanism. It  predicts that the
network always emerges (i.e for $t> 10t_c$, say)
from the phase transition epoch, almost instantaneously with respect to
 cosmological time, with a scale $L> (\sigma)^{1/2}/\eta$
and velocity $v < \sigma^{-1/2}$ essentially independently of
the values at formation. The drastic increase in string
 network correlation length imples that the string length density
 has diminished by a
large factor ($ 1/\sigma $) during a time in which the scale factor is
essentially constant. This string loss is due to frictional damping.

 This generic behaviour we have established qualitatively changes
the prevalent picture of network  evolution making it
consistent with the network at formation. This has drastic
implications for Baryogenesis mechanisms that rely on the high
initial length density predicted by the Kibble mechanism
\cite{vach,soni,brand}. The length density of string
available is reduced by a factor of $1/\sigma $ (i.e  $10^{-16}$)
for strings formed close to the Electroweak transition
temperature.  A higher scale  phase
transition  does not help since the fast evolution of
L in the extended Kibble regime that will occur before the
Electroweak transition cuts down the network density by an even
greater factor.

 It has been  proposed  
 that a topological gauge
string network formed above the EW transition at a temperature
 $\sim 1$ TeV could trap Z flux following the EW transition .  
For such scenarios, amongst others, the present authors
\cite{soni,nagaconf} have considered
the generation of the baryon asymmetry from the string network as given
by the Kibble mechanism.
Sphalerons \cite{sph} have a substantial attractive interaction
with  Z flux tubes as observed in Ref. \cite{soni} , where it was shown
that this flux ( in particular for double flux quantum strings )
may even be sufficient to bind sphalerons on strings.These sphaleron
beaded strings then lose string length into  loops. These loops decay
and in the  process release the bound sphalerons.

  The sphaleron is a
saddlepoint configuration at the top of the instanton potential
barrier separating vacua with different winding numbers. It can
thus fall into either adjoining vacuum but the CP violation in
the Electroweak mass matrix (below the Electroweak symmetry
breaking transition temperature) can bias the decay . The net
change in winding number that results is translated by  the
 anomaly in the baryon current  into 
a non zero Baryon Asymmetry. The maximum Baryon
density one could obtain in this way can be estimated by
assuming  ($M_W$ is the Electroweak mass scale)
 that each string length of size $\sim M_W^{-1}$ beads a
sphaleron which thereafter decays (at a temperature $\sim M_W$)
to yield a baryon number with
efficiency $\epsilon_{CP}$ ( $\epsilon_{CP}$ is a dimensionless 
parameter
characterizing the CP violation that biases the sphaleron decay).
  One obtains   ( where $n_L(M_W)$ stands for
the string length density and $L_W$ for the network scale at $T\sim 
M_W$)

\begin{equation}
{n_B\over n_{\gamma}}\sim \epsilon_{CP}  m_W^{-2} \times
n_L(M_W) \sim \epsilon_{CP} \left({T_c\over m_W}\right)^{2} 
(T_cL_W)^{-2}
\label{bdens}
\end{equation}

 In the Kibble regime the minimum value
 of $L_W$ is
$\sim {\tilde L_i} \sim \sigma^{-1/2}$ so that $n_B/n_{\gamma}
\sim \epsilon_{CP}/\sigma$. Due to this dilution of the string length
density the Baryon asymmetry produced is thus completely negligible.

 A similar  mechanism is operative for the twisted string
 configurations found in ref.\cite{vach}; the sphalerons being 
substituted
by
 twists.
Furthermore , order of magnitude estimates indicate that all sphaleron
 mediated B violation occuring in
strings \cite{brand} can  only  attain a maximum efficiency less than
the above.

These considerations apply to all scenarios 
\cite{soni,vach,brand,nagaconf}
 which are based on  topological
``descendant'' strings created above the EW phase transition ,
capable of trapping EW flux during the EW phase transition. Such
mechanisms are thus unlikely to work - as they have lost most of
their string length well above the EW transition, that is before
sphalerons can bead .

 They do not however
exclude the possibility that a metastable network formed during
the EW transition itself, with sphalerons beaded on strings, may yield 
an appreciable Baryon
asymmetry . This is because of the important difference  that in this 
case
  the string network  is
created below the temperature at which sphaleron mediated processes
 in the bulk are suppressed.Therefore,
 all string length that is lost ,no matter how fast or by what 
mechanism,
 contributes to the BA.

In conclusion, we have shown that the evolution of string
networks is strongly affected by friction if their scale is well
below the Planck scale resulting in a dilution of the string
length density by a factor of $T_c/M_P$. Thus the estimation
of the amplitude of the baryon asymmetry produced by electroweak strings
should be done as  presented in this Letter.
As a result, the length density of the string networks  produced
at the electroweak scale or slightly above it  would
be too low to explain the observed magnitude of the baryon number
in our universe. Similar considerations  apply to scenarios that
rely on a dense string network to generate a seed magnetic field
in the universe \cite{bag}. In special circumstances these constraints 
may
however be evaded . Thus if the string hardly interacts
 with other particles, that is, has a very tiny $\beta$,
then the frictional force becomes less effective and sufficient
string length may survive the phase transition
to produce a significant baryon asymmetry. Alternatively if the
 parameters $k,c$ are very small then
again the network may remain sufficiently dense past the
Electroweak Ginzburg temperature to allow string based
mechanisms to work. Furthermore, since the equations (\ref{lvt})
are derived on the basis of the Nambu-Goto approximation which
neglects effect of string thickness, they may require correction during
the epoch of string formation. These issues require further study.

\vskip .3 true cm
{\bf{Acknowledgment :}}
MN acknowledges the kind hospitality of National Physical Laboratory
and Panjab University where this work was initiated.
VS acknowledges the University Grants Commision and thanks
 Panjab University and the Abdus Salam International Centre for
Theoretical Physics for hospitality.


\end{document}